\documentclass[conference]{IEEEtran}
\usepackage{cite}
\usepackage{amsmath,amssymb,amsfonts}
\usepackage{algpseudocode}
\usepackage{algorithm,algorithmicx}
\usepackage{graphicx}
\usepackage{textcomp}
\def\BibTeX{{\rm B\kern-.05em{\sc i\kern-.025em b}\kern-.08em
    T\kern-.1667em\lower.7ex\hbox{E}\kern-.125emX}}

\usepackage{mathptmx} 
\usepackage{comment}
\usepackage{bm}
\usepackage[utf8]{inputenc}
\usepackage[hyphens]{url}
\usepackage[normalem]{ulem}
\usepackage[bookmarks=false,breaklinks=true,colorlinks,linkcolor=black,citecolor=blue,urlcolor=black]{hyperref}

\newcommand{\inpim}{{\em Ideal Non-PIM}\xspace}

\newcommand{\name}{{ESPIM}\xspace}

\newcommand{\iespim}{{\em Ideal ESPIM}\xspace} 

\newcommand{\bfloat}{{\tt bfloat16}\xspace}

\title{Efficient Sparse Processing-in-Memory  Architecture (ESPIM) for Machine Learning Inference}
\author{\IEEEauthorblockN{Mingxuan He}
\IEEEauthorblockA{\textit{Electrical and Computer Engineering} \\
\textit{Purdue University}\\
West Lafayette, IN, U.S.A. \\
he238@purdue.edu}
\and
\IEEEauthorblockN{Mithuna Thottethodi}
\IEEEauthorblockA{\textit{Electrical and Computer Engineering} \\
\textit{Purdue University}\\
West Lafayette, IN, U.S.A. \\
mithuna@purdue.edu}
\and
\IEEEauthorblockN{T. N. Vijaykumar}
\IEEEauthorblockA{\textit{Electrical and Computer Engineering} \\
\textit{Purdue University}\\
West Lafayette, IN, U.S.A. \\
vijay@ecn.purdue.edu}
}

\usepackage{graphicx}
\usepackage{ifthen}
\usepackage{calc}
\usepackage{pifont}
\usepackage{soul}
\usepackage{fancyhdr}
\usepackage{xspace}
\usepackage{multirow}
\usepackage[table]{xcolor} 
\definecolor{lightgray}{gray}{0.9}
\newenvironment{stripetabular}{\rowcolors{2}{white}{lightgray}\tabular}{\endtabular}

\newcounter{hours}
\newcounter{minutes}

\newboolean{timeofmake}
\setboolean{timeofmake}{false}

\newcommand{\ignore}[1]{}

\newcommand{\dontinclude}[1]{ }

\newcommand{\putsec}[2]{
\section{#2}\label{sec:#1}
}

\newcommand{\putsubsec}[2]{
\subsection{#2}\label{sec:#1}
}

\newcommand{\tabput}[3]{
\begin{table}[t]
\caption{#3 \label{tab:#1}}
\vspace{-0.15in}
\begin{center}
{
#2
}
\vspace{-0.20in}
\end{center}
\end{table}
}

\newcommand{\tabputW}[3]{
\begin{table*}
\caption{#3 \label{tab:#1}}
\vspace{-0.15in}
\begin{center}
{
#2
}
\end{center}
\vspace{-0.1in}
\end{table*}
}

\newcommand{\figput}[4][1.0\linewidth]{
\begin{figure}[t]
\begin{minipage}{\linewidth}
\footnotesize 
\begin{center}
\includegraphics[width=#1]{figures/#2}
\end{center}
\vspace{-0.15in}
\caption{#4 \label{fig:#2}}
\end{minipage}
\vspace{-0.15in}
\end{figure}
}

\newcommand{\figputW}[4][\linewidth]{
\begin{figure*}
\begin{minipage}{\linewidth}
\footnotesize 
\begin{center}
\includegraphics[width=#1]{figures/#2}
\end{center}
\vspace{-0.2in}
\caption{#4 \label{fig:#2}}
\end{minipage}
\end{figure*}
}

\newcommand{\figref}[1]{Figure~\ref{fig:#1}}
\newcommand{\tabref}[1]{Table~\ref{tab:#1}}

\newcommand{\secref}[1]{Section~\ref{sec:#1}}

\begin{document}
\maketitle
\thispagestyle{plain}
\pagestyle{plain}
\begin{abstract} 
Emerging machine learning (ML) models (e.g., transformers) 
involve memory pin bandwidth-bound matrix-vector (MV) computation in inference. By avoiding pin crossings, processing in memory (PIM) can improve performance and energy for pin-bound workloads, as evidenced by recent commercial efforts in (digital) PIM. However, PIM imposes stringent area and  energy constraints. 
Sparse models can improve performance and energy of inference without losing much accuracy. Further, unstructured sparsity is higher than structured sparsity for similar or better accuracy.
Thus, our target is unstructured, one-sided, weight-only sparsity where the vector is dense due to little use of ReLu in the models.
However, unstructured sparse inference injects the key challenges of uncertainty, irregularity, and load imbalance into a dense PIM's synchronous operation across all the banks which reads the matrix cells from each bank and broadcasts the vector elements to all the banks exploiting DRAM organization. 
To address these challenges efficiently while staying within PIM's constraints, we propose \name which makes four contributions: (1) Because matrix sparsity 
increases the  vector broadcast bandwidth demand per matrix column-read, 
\name employs a {\em fine-grained interleaving} of the matrix cells 
so that each vector broadcast is  shared among multiple rows in each bank, cutting the bandwidth demand. 
(2) As a {\em headless} architecture, \name mostly avoids on-chip control's area and energy despite sparsity's uncertainties by exploiting the observation that the sparsity is data-dependent  but  static and known before inference. Accordingly, \name employs {\em static data-dependent scheduling (SDDS)} to derive the sparse MV's cycle-level schedule and to insert the appropriate stalls for correctness. (3) Because a matrix cell's matching vector element may be broadcast much later than the cell's column-read, \name {\em decouples the matrix cell values and their indices}, placing the indices  ahead of the values to enable  prefetching of the vector elements. We extend SDDS for performance and correctness with the decoupled prefetching. 
(4) Finally,  we {\em simplify the switch} required to select the vector elements that match the matrix cells instead of a brute-force, impractically-large design. 
We extend SDDS to improve performance by achieving fewer conflicts in the simplified switch. In our simulations,\name 
achieves 2x average (up to 4.2x) speedup over and 34\% average 
(up to 63\%) lower energy than Newton while incurring under 5\% area.

\end{abstract}

 \putsec{intro}{Introduction}

Machine learning (ML) has emerged as a prevalent domain for visual and linguistic processing. Convolutional neural network inference involving matrix-matrix multiplication (MM) is compute-bound (O($n^3$) compute with high reuse versus O($n^2$) space for $n$x$n$ matrices). In contrast, recent
decoder-only transformer-based inference using relatively smaller models deployed in edge devices with little or no input batching involves memory pin bandwidth-bound matrix-vector (MV) multiplication (O($n^2$) compute with little matrix reuse versus O($n^2$) space).  
Such edge deployment is attractive in privacy-sensitive and wireless bandwidth-limited scenarios. For instance, companies may privately deploy large, high-accuracy models instead of sending sensitive data to the Cloud. Such deployments would not get Cloud-level request traffic or  batching. A server utilized only at 20-30\% due to low request rate  (and batching) may still be faster and more cost-effective than manual data processing. Thus, even large models may have use cases with low batching (i.e.,  low weight matrix reuse).
Memory pin-bandwidth boundedness due to high spatial locality but poor reuse is {\em different}  from general memory bandwidth boundedness due to poor spatial {\em and} temporal locality  leading to DRAM row misses so that {\em all} the banks are busy and are the bottleneck (i.e., {\em not} pin-bound). 

Processing in memory (PIM)~\cite{Stone-PIM,Kogge-PIM,IBM-AMC, IRAM, Kogge-DIVA, flexram,mcdram} is a promising approach for  pin-bound workloads. PIM places compute units within DRAM to exploit the high internal bandwidth of DRAM banks, which far exceeds the DRAM pin bandwidth, and avoids  off-chip movement of DRAM data (e.g., 16 banks provide 16x speedup opportunity and significant energy reduction over non-PIM systems). Though known for decades, PIM has not been adopted mainly due to the lack of compelling workloads like MV-based ML models. Indeed, Samsung's Function In Memory (FIM)~\cite{FIM,FIM-ISSSC} and Hynix's Accelerator in Memory  (AiM), called Newton~\cite{Newton,Newton-ISSSC}, point to significant commercial interest. Our focus is digital PIM; not analog PIM~\cite{pim-Puma,pim-Pipelayer,pim-ISAAC,pim-Prime} which faces well-known circuit issues.

PIM provides high bandwidth but limits area and hardware complexity (for logic in DRAM process). Accordingly, Newton employs a {\em headless} architecture which places {\em only} the datapath in the DRAM whereas the host provides the  control via read/write-like commands (no instruction pipeline,  register file, or caches). The multiply-accumulate units (MACs) and a few buffers {\em alone} add around 25\% area to Newton~\cite{Newton}.  For
more generality, FIM adds instruction processing, a register file, and a load/store unit, but incurs around 50\% area  as evidenced by its half the normal capacity~\cite{FIM}. 

Sparsity -- zeros in operands  --  can improve speed and energy in inference by reducing the work. Pruning followed by retraining creates sparse models that are nearly as accurate as the dense models~\cite{pruning, pruning1}. Structured sparsity can reduce hardware complexity~\cite{wen2016-ssl,STC} (e.g., Ampere's 2:4 sparsity at 50\%). Even small models appropriate for edge deployment can be pruned without losing much accuracy (e.g., RoBERTa's~\cite{roberta} authors achieve 80\% structured sparsity for  LLaMA-7B~\cite{sheared} and another study achieves 50\%~\cite{sun2024a}). However, structured sparsity is lower than unstructured sparsity (80-90\%) for similar or better accuracy~\cite{sparsezoo,cerebras,sparsegpt}. 
Consequently, we focus on unstructured sparsity though our approach applies to structured sparsity. In sparse MV, the weight matrix is  sparse whereas the vector is dense due to almost no use of ReLU in transformers. Thus, our target is unstructured, one-sided, weight-only sparsity where the vector is  dense. While dense PIM~\cite{Newton, FIM, FIM-ISSSC, Newton-ISSSC}  and sparse non-PIM~\cite{cnvlutin,  scnn, sparten, dstc,cambricon,cambriconx, s2ta} ML accelerators have been explored extensively, sparse PIM is less-explored. SpaceA~\cite{spacea}, a sparse PIM, targets hyper-sparse MV in High Performance Computing (HPC) with  99.9--99.999\% sparsities in large matrices (e.g., $10^5$x$10^5$). However, SpaceA incurs complexity and overheads for sparse ML models whose sparsities are considerably different than HPC's (e.g., 80-90\%), As such, SpaceA is not a good fit for sparse ML, as our results confirm. 

Sparsity introduces the significant challenges of  uncertainty, irregularity, and load imbalance to dense PIMs like Newton. 

Exploiting DRAM's internal buses, Newton broadcasts a vector {\em slice} (e.g., 16 elements) to all the banks which column-read the matrix data in parallel to each broadcast. 
The banks then  compute {\em in lockstep} the MV partial product for their respective matrix rows.
This lockstep operation is key to keeping both the off-chip  command and on-chip vector bandwidth demands feasible while mostly avoiding  on-chip (global or per-bank) control. Because holding a {\em vector-row} -- a DRAM row-sized sub-vector --  at each bank incurs a high area overhead, the vector-row is broadcast, one slice at a time, to  all the banks for every DRAM  row of the matrix (otherwise, the broadcast buses would idle during the banks' column-reads).
To achieve vector reuse, the matrix uses DRAM row-wide {\em coarse-grained interleaving} so Newton marches down each bank's matrix DRAM rows for the same vector-row.

\name adopts Newton's headless architecture and addresses the   challenges sparsity poses for PIM. 
The root issue is that while the vector is dense the matrix is sparse and compressed,
so any DRAM column-read of the sparse matrix corresponds to dense vector elements spread over 
multiple columns (e.g., 90\% sparsity means 16 sparse matrix cells  span 160 dense vector elements, on average). 
While  the banks  operate in lockstep by receiving the vector broadcasts, the non-zero cell indices in each bank are {\em different}. Further, every vector element has a high probability of being used in some bank so no element can be removed from the broadcast (e.g., even at 90\% sparsity, this probability is more than 81.4\% for 16 banks). 
Unfortunately, the broadcast bandwidth cannot support (1) individual vector transfers to each bank, or (2) ultra-wide transfers (e.g., 160-element).  
Therefore, despite the sparsity, \name continues to broadcast sequentially the slices in a vector-row, so that each bank selects the vector elements relevant for each column-read. However, four issues remain which \name addresses {\em efficiently while staying within PIM"s constraints}. 

First, Newton rate-matches each DRAM column-read of the matrix with exactly one vector slice broadcast for computing {\em one} partial inner product per bank. However, this schedule poses the problem that at 90\% sparsity,
each column-read would require 10 times more vector slice broadcasts as Newton, eliminating any sparsity advantages.  
Instead of placing a sparse matrix row along a DRAM row, we place along a DRAM row  the first
element of each of $k$ consecutive sparse matrix rows and then the next element and so on (e.g., $k$  = 16). 
In this new, {\em fine-grained interleaving} for {\em sparse reuse}, different from Newton's coarse-grained interleaving for {\em dense reuse}, $k$ consecutive sparse matrix rows reuse each vector broadcast.
Thus, a bank's MACs compute $k$  partial inner products instead of just one as in Newton. Crucially, the new layout achieves $k$-times  fewer vector broadcasts,
restoring sparsity advantage, at the modest cost of a $k$-element output vector per bank instead of one scalar (an $k$-element output vector would  not improve Newton which does not require more vector broadcasts). This fine-grained interleaving fundamentally enables \name to continue to exploit
vector broadcasts. Because of around 31\% bandwidth overhead for sparse representation (only 11 matrix elements fit in a column), the maximum speedup over Newton for 90\% sparsity is 0.69 * 10 = 6.9. 

Second, a given vector slice broadcast may have matching elements for more than one DRAM column-read in some but not all of the banks, requiring {\em for correctness} a {\em data-dependent} stall of the  next broadcast (and dummy matrix cells) until the current slice is consumed fully. However,  
there is little on-chip control -- global or per-bank --  to handle such uncertainty. Fortunately, the sparsity is data-dependent on the specific matrix but is static and known offline at training. Accordingly, we propose {\em static data-dependent scheduling (SDDS)} for correctness by deriving the full cycle-level schedule of the sparse MV computation via cycle-accurate simulations -- once, at training.

Third, because the vector slices within the vector-row are broadcast 
sequentially, a given matrix cell may be stalled for a later vector element.  To alleviate such stalls, we propose to 
{\em decouple the matrix cell values and indices} by placing the indices well ahead of the corresponding matrix cells in the DRAM layout, enabling the matching vector elements to be prefetched. To this end, \name employs two  {\em non-search}, strict  FIFOs per MAC, a {\em matrix cell-index FIFO (iFIFO)} and a {\em vector-element FIFO (eFIFO)} (e.g., 8 entries each).
The iFIFO holds the prefetched indices from the DRAM column-reads to insert  into the eFIFO the relevant vector elements from each broadcast. Despite the decoupling, the banks continue synchronous operations.
We extend SDDS to include the decoupled prefetching for performance and to stall the broadcasts (and to insert dummy matrix cells) upon full or empty FIFOs for correctness.
While SpaceA prefetches the vector from a CAM into its load queue to be searched, \name's  continues to extract the matching prefetched vector elements on the fly from the  broadcasts into  the simple FIFOs.

Finally, sparsity destroys the one-to-one correspondence in Newton between the vector elements in a broadcast and the matrix cells in a DRAM column-read.  From a vector slice broadcast in \name, each MAC in a bank has to select the element  corresponding to the MAC's  matrix cell index. 
However, brute-force design  would lead to an impractically large switch (e.g., a 16x11 switch for 16 elements and 11 MACs at each bank). We {\em simplify  the switch} by exploiting the $t_{CCD}$-constrained time between broadcasts (i.e., use a 4x11 switch, made of 11 4-to-1 multiplexers, sequentially 4 times). We extend SDDS to improve performance by achieving fewer conflicts in the simplified switch.

Further, \name adopts SparTen's greedy load balancing~\cite{sparten}  dense and sparse matrix rows in different banks. Our simulations show that \name achieves 2x average (up to 4.2x) speedup over and 34\% average (up to 63\%) lower energy than Newton while incurring under 5\% area.

\putsec{background}{Background and challenges}

Recall from~\secref{intro} that the key performance and energy advantages of PIM come from exploiting the high internal DRAM bandwidth of multiple banks whose data would be serialized in conventional DRAM through narrow pins (e.g., 16 banks have 16x higher internal bandwidth than a conventional DRAM). High-bandwidth DRAM exploit wider paths than conventional pins via 3-D or 2.5-D interconnection between the CPU and memory (e.g., HBM~\cite{hbm2e}). However, the conventional DRAM's internal bandwidth is typically higher than HBM's external bandwidth. Of course, PIM can exploit HBM's even higher internal bandwidth as well. 

\putsubsec{pim}{Dense PIM}

\figput{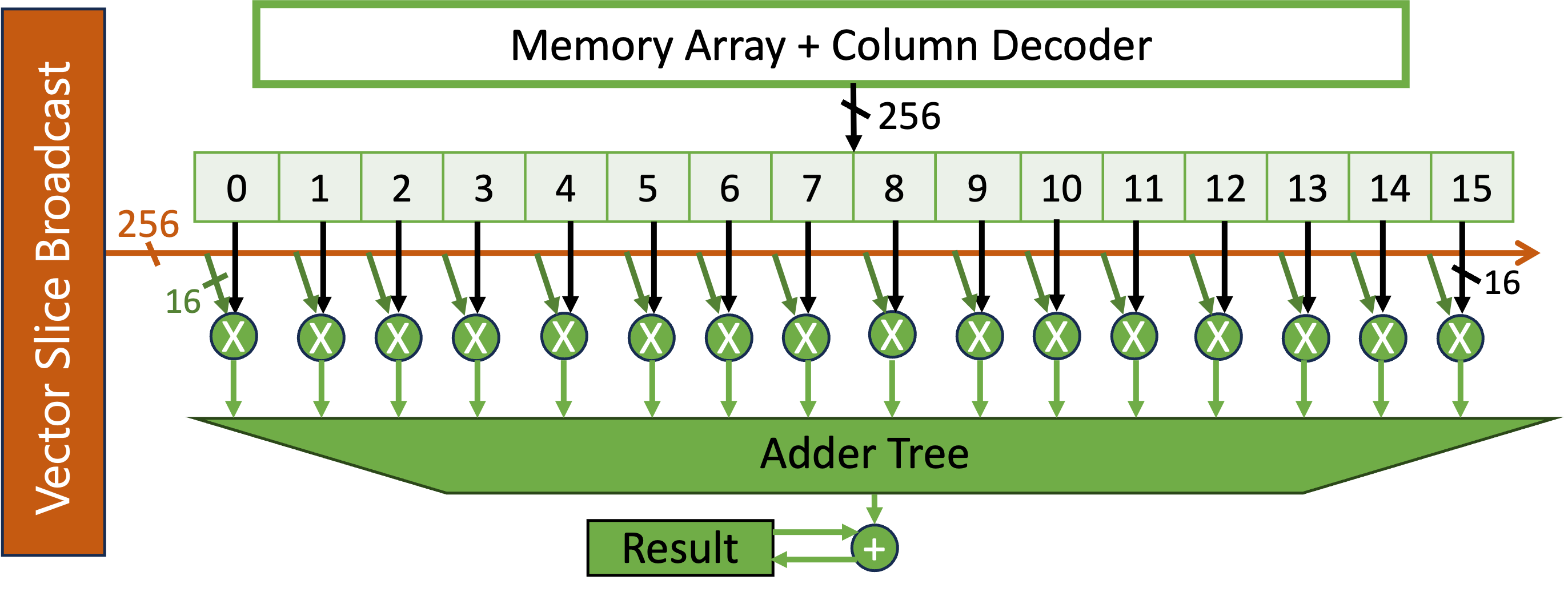}{}{Newton's datapath for one bank}

As discussed in~\secref{intro}, to exploit PIM's bandwidth advantage within its area and hardware complexity constraints (e.g., no on-chip inter-bank communication), Newton~\cite{Newton,Newton-ISSSC} employs a {\em headless} architecture where only the datapath is in the DRAM whereas the host provides the control  via read/write-like commands. 
The filter matrix is held in the DRAM and the vector is sent from the host to the PIM which holds a vector-row in the {\em global buffer} common to all the banks in  the channel.  Newton exploits DRAM's internal buses  to broadcast a vector slice  from the global buffer to all the banks which latch the slice (\figref{Newton_datapath.png}). In parallel to each broadcast, the banks column-read the matrix data. The banks
synchronously compute their respective partial products, conserving the off-chip command and on-chip vector bandwidths {\em even} without much on-chip control. To avoid per-bank vector-row area overhead, each slice of the  vector-row is broadcast for every DRAM row of the matrix (per-bank area versus common broadcast energy trade-off). Without the repeated broadcasts, the broadcast buses would idle anyway during the column-reads of the banks. 
Each bank computes its partial {\em inner product} producing a scalar result per  vector-row (\figref{Newton_datapath.png}).  Newton achieves vector reuse by marching down each bank's DRAM rows for the same  vector-row where the matrix uses DRAM row-wide {\em coarse-grained interleaved} layout (\figref{interleaving1.png}). In the figure, (1) the numbering shows how the matrix is linearized in memory, and (2) the color coding shows the corresponding matrix cells and vector elements.

After a matrix DRAM row is exhausted while accumulating the partial product in the scalar result, the host reads out all the bank's results (e.g., 16 scalars from 16 banks). Marching down the bank for each vector slice instead of vector-row would avoid the repeated vector broadcasts but would incur repeated read-out of the partial products.  
The vector slice broadcast efficiently captures  reuse across the banks. Further, one transfer  of the vector-row from the host to the DRAM is reused numerous times across all banks and their rows. Such reuse is key to  conserving the host-DRAM bandwidth, An implication of  PIM's constraints is fewer compute units than a GPU or TPU (hundreds versus thousands) so that compute-bound workloads (MM or batched MV) would likely be slower in all PIM (not only Newton).

\figput{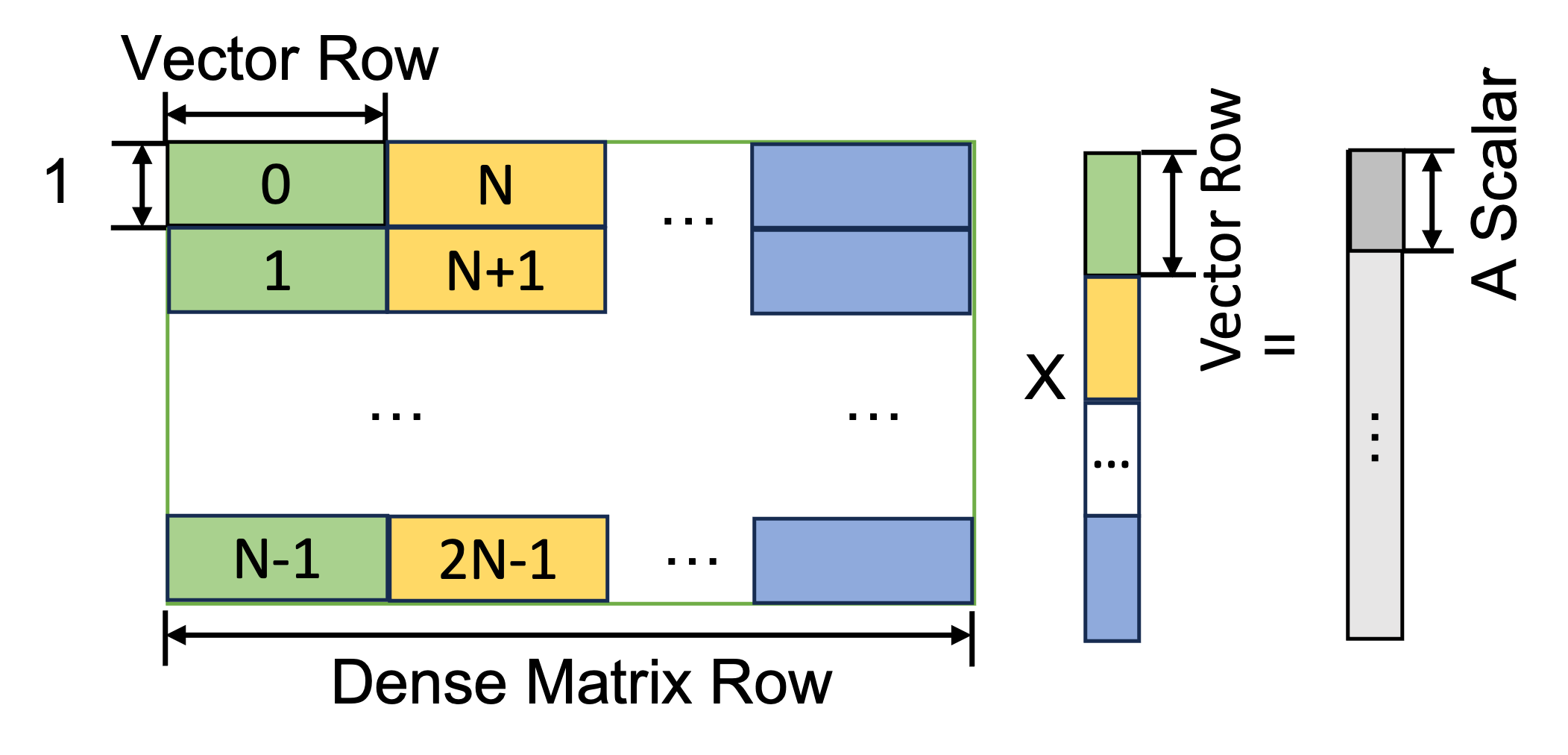}{}{Coarse-grained interleaving in dense matrix for one bank}

The loading of the  vector-row into the global buffer is amortized over all the DRAM rows of all the  banks. For a vector-row, a matrix DRAM row is activated in each bank followed by all the column-reads and multiplication of the row, and the result read out (\figref{Newton_operation.png}). 
While conventional DRAM row activation is limited by $t_{FAW}$ constraints due to power, PIM's compute power for all the MACs in parallel far exceeds that of all-bank activation~\cite{FIM-ISSSC}. As such, power delivery for the MACs can also cover all-bank activations, which occur necessarily {\em before} MAC operation,  eliminating the $t_{FAW}$ constraint. Therefore,  Newton's 
(overhead of) sequential activations of groups of four banks can be replaced by (the smaller overhead of) all-bank activations. In Newton, the sequential activation overhead is the main reason for deviating from the ideal speedup of the number of banks.

\putsubsec{challenges}{Sparsity challenges in PIM}
Recall from~\secref{intro} that we focus on unstructured, one-sided weight-only sparsity where the vector is dense.  Sparsity introduces uncertainty, irregularity, and load imbalance into dense PIM's above schedule. First, at 90\% sparsity (structured or unstructured), a column-read of the sparse matrix  needs around 10 dense vector slice broadcasts to find matching vector elements (e.g., the matching vector elements for 16 matrix cells span 160 dense vector elements on average). 
In Newton, however,  the column-reads and the slice broadcasts are rate-matched. Increasing the broadcast bandwidth demand by 10x is impractical.  
Second,  a vector slice broadcast may have matching elements for more than one DRAM column-read in some but not all of the banks.
The next broadcast cannot occur until the current slice is consumed fully, 
requiring a {\em dynamic} stall  of the vector broadcast to allow later DRAM column-reads to consume fully the current vector slice -- a correctness requirement. 
However, given the headless nature of the architecture there is little on-chip control -- global or per-bank --  to handle such dynamic conditions that vary across the banks.
Third, because the matching vector elements may span many vector slice broadcasts which occur sequentially, a given matrix column-read may have to wait for some future broadcast. Such waiting injects significant latencies into the PIM operation. 
Finally, because sparsity destroys the one-to-one correspondence  between the vector elements in a broadcast and the matrix cells in a DRAM column-read, a switch is needed to select
the relevant elements from the broadcasts.  Assuming 16 elements in a  broadcast and 11 MACs per bank (16*16 bits = 256 bits of broadcast width),  any of the 16 elements in a broadcast  may match any of the matrix cells. As such, a naive design may use an impractically large  16x11 switch. Because each bank's non-zero cell indices are different irrespective of Ampere-like structured or unstructured sparsity, a switch may be unavoidable for either type of sparsity.

\figput{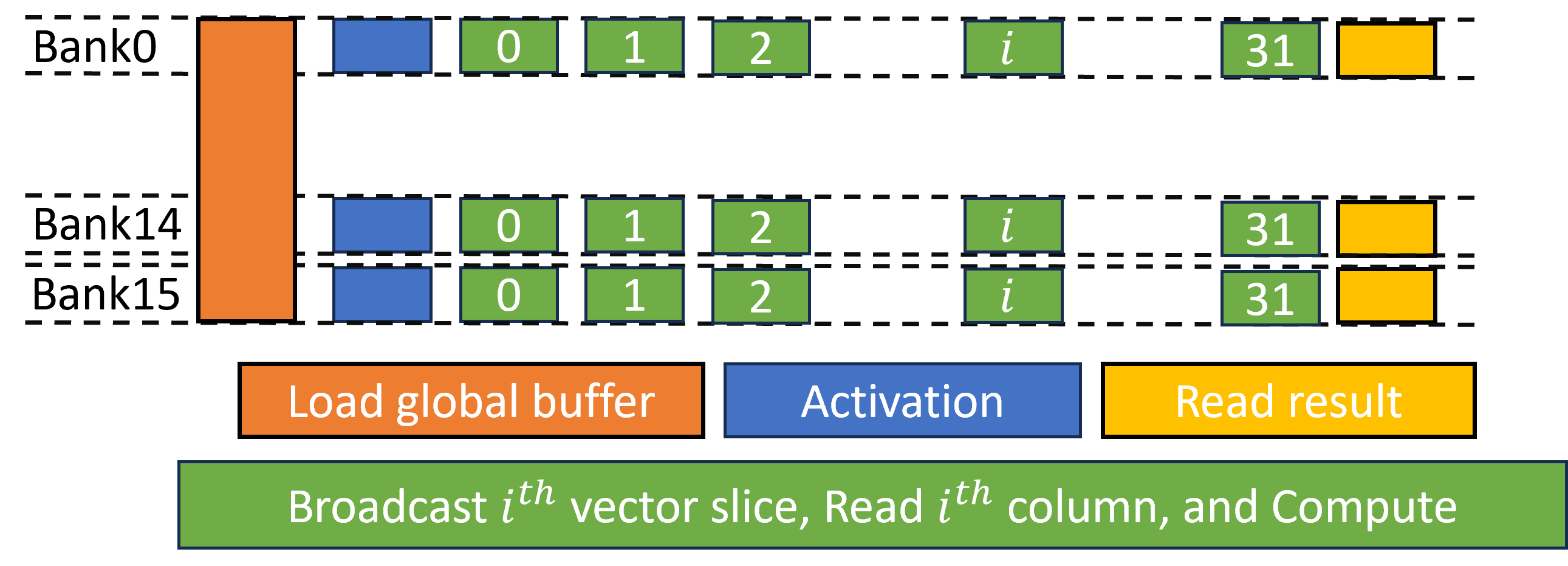}{}{Newton's operation across all banks}

For reference, SpaceA~\cite{spacea} takes a
hardware-intensive approach to target hyper sparsity. 
SpaceA employs a per-bank CAM to provide the vector to the MACs
instead of exploiting DRAM's organization to broadcast the vector.
SpaceA employs a scratchpad to cache the matrix data instead of using the bank's row buffer. To handle the uncertainty of sparsity, SpaceA employs on-chip, per-bank control. To extract the  
vector elements matching the matrix cells, SpaceA
employs two-level, associatively-searched load queues.

\putsec{AiMarch}{\name} 

Recall from~\secref{intro} that \name addresses the above challenges via four contributions. (1) To avoid 10x more vector broadcasts, \name adopts a {\em fine-grained interleaved layout}
where a bank's MACs compute $n$  partial inner products per bank  (e.g., 16) instead of just one as in Newton so that each vector broadcast is used by $n$ consecutive matrix rows achieving $n$-times  fewer vector broadcasts. (2) The dynamic uncertainty of extracting varying numbers of matching  elements from vector slice broadcasts across the banks requires broadcast stalls until
all the matches of the current slice are extracted -- a correctness issue.
To handle this uncertainty with little on-chip control in \name's headless architecture, \name exploits the observation that though data-dependent on the specific matrix, the sparsity is static and known at training. Accordingly, \name proposes  {\em static data-dependent scheduling (SDDS)} for correctness by deriving the full cycle-level schedule of the sparse MV computation  via cycle-level simulation so that the host's command sequence is correct.
(3) To address the latency of sequential vector slice broadcasts within a vector-row out of which the matching vector elements are selected, \name proposes {\em to decouple the matrix cell values and indices} by
placing the indices well ahead of the corresponding matrix cells in the DRAM layout, enabling  the indices and vector elements to be prefetched.  
We extend SDDS to achieve the decoupled prefetching for performance and to stall the broadcasts (and to insert dummy matrix cells) for correctness upon the FIFOs being full or empty. (4) Finally,  we {\em simplify the switch} needed to extract the matching vector elements  from each broadcast 
by serializing the wide selection into multiple sequential narrower selections in  the $t_{CCD}$-constrained time between broadcasts. 
We extend SDDS to improve performance by reducing the number of conflicts due to the simplified switch.

\putsubsec{naive}{Naive operation overview}
Following Newton, \name broadcasts the dense vector slices to 
the banks which column-read the sparse matrix data in parallel. 
Though all the banks receive the vector broadcasts and operate in lockstep, each bank's non-zero indices are {\em different}.Further, no vector element can be removed from the broadcast  as every  element has a high probability of being used in some bank (e.g., even at sparsity  as high as 90\%, every element has more than 81.4\% chance of being used in at least one of 16 banks). 
Unfortunately, the broadcast bandwidth cannot support (1) individual vector slice transfer to each bank, or (2) ultra-wide broadcasts (e.g., 160-element).  Consequently, \name follows Newton to continue to broadcast the  vector-row, advancing sequentially one slice at a time, to all the banks.
Each bank selects the vector elements relevant for each column-read.

\figput{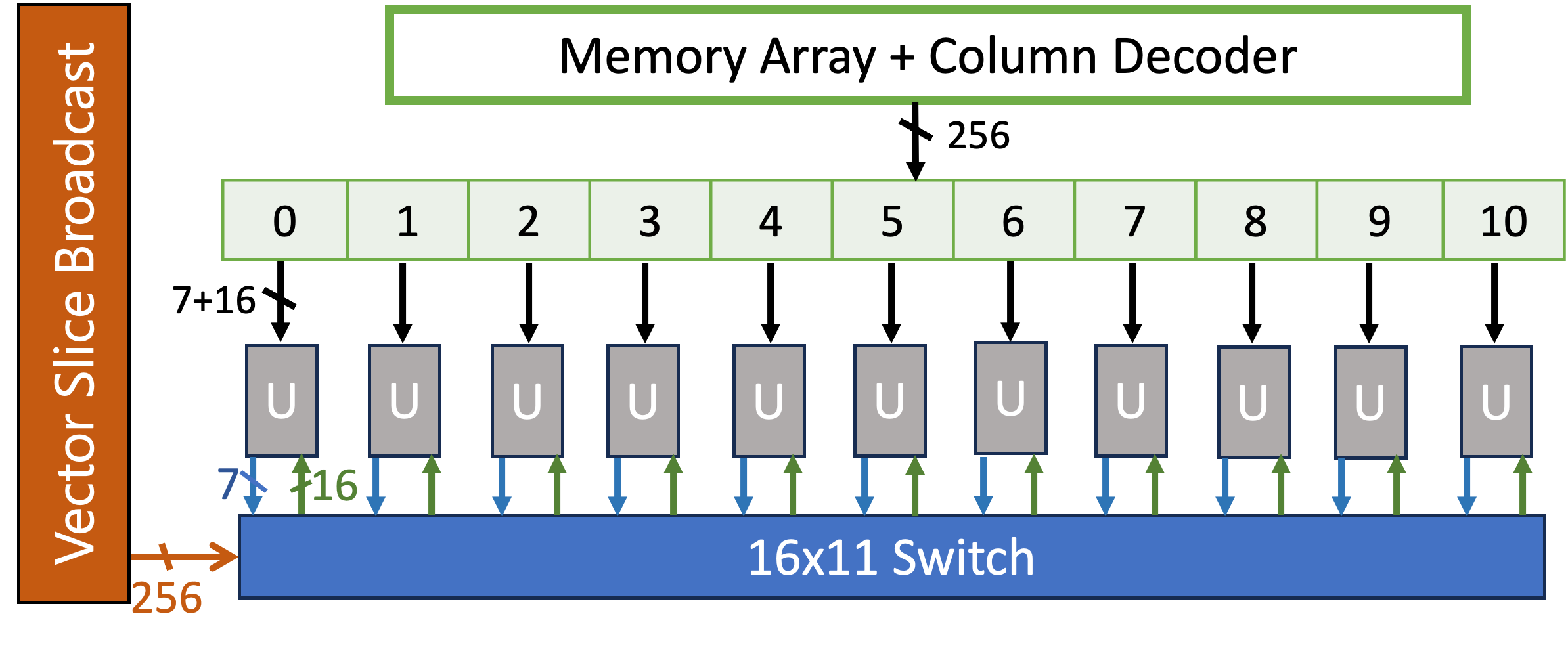}{}{\name unoptimized sparse-only datapath ($U$ is an execution unit comprising a MAC and other components)}

The broadcast is 256 bits wide, providing 16 16-bit elements which are latched in each bank. Each matrix
data column-read has both the values and indices of $k$  non-zero cells corresponding to $k$ MACs in each bank; the indices are in increasing order.  
Even though the column-read width is also 256 bits, $k$ = 11 due to the sparse index overhead.
Based on the indices, a 16x11 switch extracts the elements from the vector broadcast latch that match the matrix cells in the column-read (\figref{datapath.png}). This figure shows a datapath only for sparse models. We later extend the datapath to support flexibly  both dense and sparse models (\secref{flexible}). The execution units ($U$ in the figure) in each bank compute the partial inner product of the  matching elements in the vector-row and matrix cells. bank's 
scalar partial product is read out to the host at the end of each DRAM row.  As in Newton, a vector-row is  held in the global buffer common to the entire channel and reused by marching down the matrix DRAM rows without requiring the vector-rows to be sent repeatedly from the host to the PIM. In the following sections, we modify this naive operation to incorporate \name's  optimizations.

\putsubsec{layout}{Fine-grained interleaved layout}

In the above naive operation, because the vector is dense and the matrix is sparse and compressed, a sparsity of  90\% means a matrix column-read spans $10$ vector slices requiring $10$ broadcasts, on average. This requirement would eliminate any sparsity advantage. 
To address this issue, we  propose a {\em fine-grained interleaved} layout
where instead of placing a sparse matrix row  along a DRAM row, we place along a DRAM row  the first element of each of $k$ consecutive sparse matrix rows and then the next element and so on (\figref{interleaving2.png}).  
This layout targets sparse reuse whereas Newton's coarse-grained interleaving achieves dense reuse. 
In this layout, where $k$ consecutive sparse matrix rows reuse the vector elements, a bank's $k$ MACs compute $k$  partial inner products per bank instead of just one per bank as in Newton (i.e., each matrix row is mapped to a MAC). Thus, the new layout achieves $k$-times  fewer vector broadcasts, restoring sparsity advantage. 
This layout comes at the modest cost of a $k$-element output vector per bank instead of one scalar. We note that an $k$-element output vector would  not improve Newton which does not require more vector broadcasts (so the extra output buffering would be an unnecessary overhead). Further, the output read bandwidth remains the same as Newton's whose host reads an output scalar per bank for each matrix row whereas \name's host reads an output $k$-element vector per bank for every $k$ matrix rows.

\figput{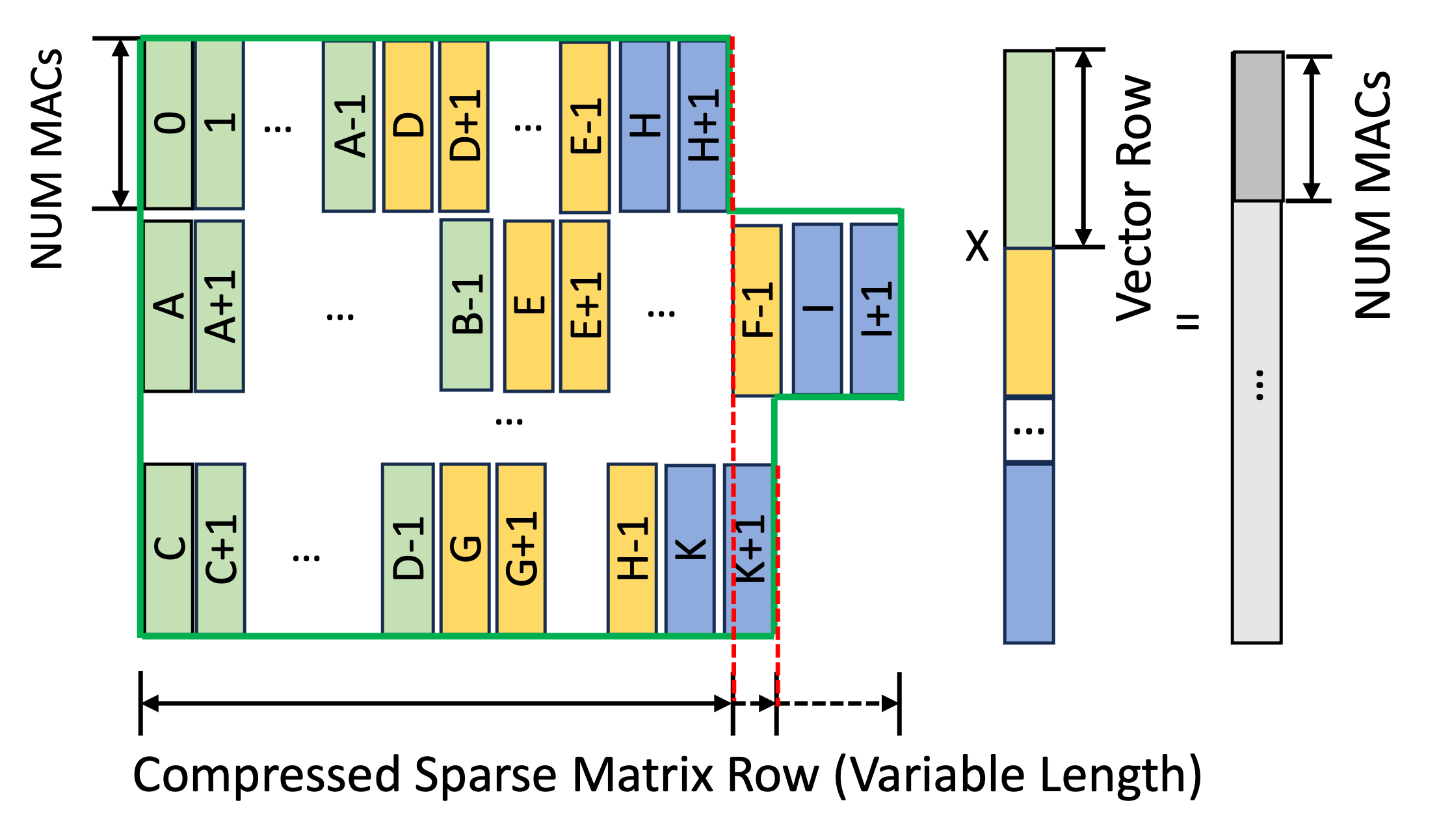}{}{Fine-grained interleaving in compressed sparse matrix for one bank}

In~\figref{interleaving2.png}, (1) the numbering shows how the matrix is linearized in memory, and (2) the color coding shows the corresponding matrix cells and vector elements. Each {\em matrix row segment} ends at the rightmost matrix cell that falls within the corresponding vector-row (e.g., 0 to A-1).  
Because of the sparsity, the physical length of 
each matrix row segment may be different. However, each segment end is known statically, at training, so the matrix can be linearized as shown. 
Further, a matrix row segment may span more than one DRAM row and may end in the middle of a DRAM row. Nevertheless, only one MAC computes the inner product for the entire  segment. Thus, all the cells  of a segment  -- irrespective of their DRAM row -- contribute to the same inner product accumulated by the corresponding MAC. 

A subtle point is that with this layout each MAC receives 1.6 vector elements on average assuming 16 elements per broadcast and 90\% sparsity, compared to Newton where each MAC receives exactly 1 element per broadcast. Thus, \name is bank
bandwidth-bound (the MACs are rate-matched to the bank) with some
surplus vector broadcast bandwidth which is consumed by broadcast stalls due to the irregularity of the matrix's sparsity. As sparsity increases the surplus  decreases and at high sparsities the surplus turns into deficit (e.g., at 95\% sparsity each MAC receives 0.8 elements per broadcast). Conversely, at lower sparsities the surplus grows (e.g., at 0\% sparsity each MAC receives 16 elements per broadcast at the cost of extra output buffering without any benefit). 

This fine-grained interleaving fundamentally enables \name to continue to exploit vector broadcasts.The need for extra vector broadcasts is independent of structured or unstructured sparsity, though  Ampere-like structured sparsity's demand may be lower than that of unstructured sparsity due to lower sparsity.
Our fine-grained interleaved layout applies to both sparsity types.

\putsubsec{representation}{Sparse representation}

Instead of providing a long index within the entire sparse matrix row,  each non-zero  matrix cell provides its position within the corresponding vector slice of 16 elements requiring only  4 bits of {\em index}. 
Because the matching vector element for a given matrix cell may be in a future  vector slice and not in the current  slice, we add a {\em valid bit} to each cell. While an invalid cell's index and value bits  are wasted, the probability that a given matrix cell does not match any of the 16 elements in a slice is low even for 90\% sparsity.
Moreover, we later remove the dummy values for most invalid cells. While a dummy value of zero can be used instead of the valid bit, we wish to avoid the energy of multiplying by zeros or of zero detection when the value is not zero.

With this sparse representation, only 11 matrix cells (FP16 values + 7 metadata bits each where the last two metadata bits will be added later) can fit in a column read (of 256 bits). Accordingly, each bank has only 11 MACs instead of Newton's 16, implying that the maximum speedup over Newton for 90\% sparsity is 11/16 * 10 = 6.9. Note that placing the indices and values in separate DRAM rows is equivalent.

\putsubsec{SDDS}{Static data-dependent scheduling (SDDS)}

A given vector slice may be needed by more than one DRAM column-read (e.g., the simple case of a matrix row's consecutive cells necessarily fall into consecutive column-reads in our fine-grained interleaving but the matching vector elements may be in the same slice). Consequently, the next vector slice broadcast may need to be stalled until  later DRAM column-read(s) have consumed fully the current vector slice.
Such uncertain scenarios require a {\em data-dependent} stall of the  next vector broadcast  {\em for correctness}. However, there is little on-chip control -- global or per-bank. We exploit the fact that the sparsity is data-dependent on the specific matrix but is static and known at training. Accordingly, we propose static data-dependent scheduling (SDDS) for correctness by cycle-accurately simulating the sparse MV computation to derive the full cycle-level schedule. This simulation is done once, at training. SDDS is distinct from conventional static scheduling which is data-independent and from
inspector-executor approach which inspects the input data for every run as the data changes from run to run unlike ML filters in inference runs. 

\figput{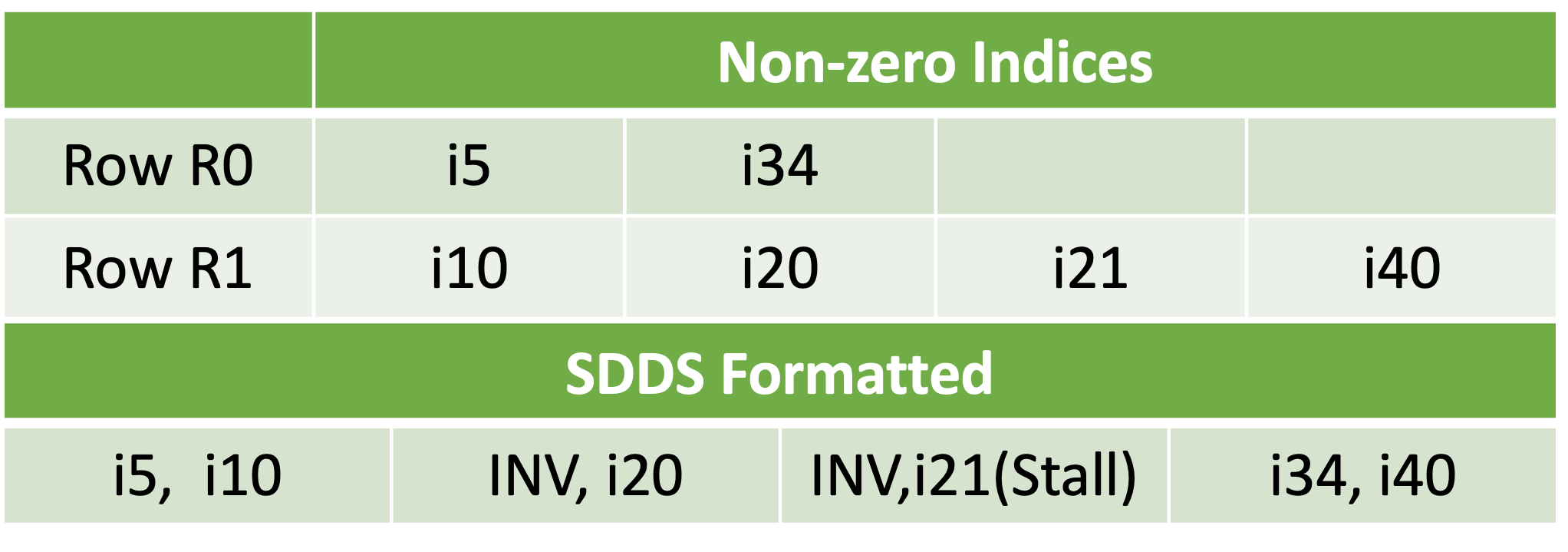}{}{SDDS example}

For each bank, SDDS builds the compressed sparse matrix from the 
uncompressed sparse matrix  in our fine-grained layout.  Starting  with the first vector slice and the uncompressed matrix,
the scheduler determines whether the next non-zero  cell in a matrix row mapped to a MAC
matches an element in the current vector slice. If so, the scheduler places the cell value and index in the  compressed matrix column position corresponding to the MAC. Otherwise, the scheduler places an invalid cell (with a dummy value) which stalls the corresponding MAC. \figref{SDDS.png} shows
the non-zero indices in two sparse matrix rows $r0$ and $r1$ mapped to  MACs $M0$ and $M1$, respectively. Assuming the
vector slice is 16 elements, SDDS packs indices $i5$ and $i10$ in the matrix column positions for MACs $M0$ and $M1$, respectively, in accordance with our fine-grained interleaving and schedules a broadcast of the vector slice $v0$-$v15$ (we do not show the cell values for clarity).  MAC $M0$'s next non-zero index is $i34$ which falls beyond  the next  slice $v16$-$v31$. Thus, SDDS packs an invalid index and $i20$ for MACs $M0$ and $M1$, respectively, and schedules a broadcast of the slice $v16$-$v31$. In this manner, the scheduler fills the columns of the compressed matrix across all the banks.

SDDS determines whether the current  vector slice is consumed fully across all the banks at the end of each compressed matrix  column-read (i.e., each bank's next non-zero index falls beyond the current slice). If so, the next slice broadcast is scheduled. If not, the next slice broadcast is stalled until the current slice is consumed fully by later column-reads. 
Stalling the current slice induces invalid matrix cells in the banks where the next non-zero index falls in a later  slice. In~\figref{SDDS.png}, 
while row $r0$ has no matching index for the vector slice $v16$-$v31$ ($r0$'s next non-zero index is $i34$), row $r1$'s $i20$ and $i21$ fall in that slice. Therefore, SDDS packs an invalid index and $i21$ for MACs $M0$ and $M1$, respectively, and schedules a vector broadcast stall which is a next  column-read without the accompanying next slice broadcast. Next, 
SDDS packs $i34$ and $i40$ for MACs $M0$ and $M1$, respectively, and  schedules a broadcast of the slice $v32$-$v47$.
The compressed matrix column advances once the previous column is full irrespective of whether the vector slice stalls.

SDDS formats the compressed matrix including the metadata in the fine-grained interleaved layout  and  generates the full schedule of commands from the host to \name. 
The basic command sequence is similar to Newton's: (1)  load global buffer with the vector, (2) activate row followed by (3) sequential next column-read accompanied by next vector slice broadcasts, and (4) result read out at the end of the matrix row. 
Because the host has to insert broadcast stalls in the command sequence when needed, SDDS creates a {\em command stream} for the host indicating the stalls.  

We assume that the host memory controller is prevented from reordering the DRAM commands (\name achieves full bandwidth without such reordering). DRAM refresh can be handled either before the processing of a DRAM row starts or after, by incorporating slack in the refresh timing~\cite{Newton}. Because SDDS scheduling is for operations within a DRAM row, refresh does not affect SDDS. 

We extend this basic scheduler (1) to include the decoupled prefetching for performance and to handle full prefetch FIFOs for correctness, and (2) to improve performance via fewer conflicts due to the simplified switch.

\putsubsec{decouple}{Decoupling matrix cell values and indices for prefetching}
The vector slices within the vector-row are broadcast sequentially so that a given matrix cell may be stalled (via  invalid cells) for a  later vector element broadcast. To alleviate such stalls, we propose to decouple the matrix cell values and indices by placing the indices well ahead of the corresponding values in the DRAM layout, enabling the indices and vector elements to be prefetched. To this end, \name employs two  {\em non-search} strict FIFOs per MAC, a matrix cell-index FIFO (iFIFO) and a vector-element FIFO (eFIFO) (e.g., 8 entries each).  The iFIFO holds the prefetched indices from the DRAM column-reads to insert the relevant vector elements from each broadcast into the eFIFO. \figref{Unit.png} expands the execution units ``U'' in~\figref{datapath.png}. 

To facilitate the prefetching of the indices, SDDS places multiple indices contiguously well before their corresponding matrix cells in the same order.
SDDS packs the indices in an {\em index-only column-read} and records the command in the {\em command stream} for the host. 
Despite the decoupling, the banks continue to operate synchronously.

\figput{Unit.png}{}{\name's execution unit}

The index-only column-read is in addition to the normal value-index column-read in which the values are for the  previously-prefetched indices whereas the current indices are for later values. 
Upon a normal column-read in parallel with a vector slice broadcast,
each index in the column-read is pushed into the corresponding  MAC's iFIFO at the tail (step \textcircled{1} in~\figref{Unit.png}). 
Each iFIFO provides its indices from the head to retrieve the matching vector elements from the broadcast via the switch (step \textcircled{2} in~\figref{Unit.png}). The switch inserts the elements into their respective eFIFO (step \textcircled{3} in~\figref{Unit.png}). Because
the switch operates sequentially reading only one index from the iFIFO and writing only one vector element into the eFIFO at a time as we explain later, the FIFOs remain single-ported. 
For {\em each} MAC, the  indices in the iFIFO, and therefore the matching vector elements in the eFIFO, are in the same order as the matrix cell values in consecutive normal column-reads. Therefore, a normal column-read triggers the multiplication of the values in the column-read and the vector elements at the heads of  the eFIFOs instead of the vector elements extracted from the current broadcast. The matrix values in a column-read are consumed immediately without any further buffering.

In contrast to a normal value-index column-read, an index-only column-read, which does not have an accompanying vector broadcast or compute in the MACs,
simply places the indices at the tail of  each iFIFO.
However,  the probability that each MAC would have multiple matching
elements within one vector slice for the multiple indices in the index-only column-read is quite low for high sparsities, forcing many invalid cells and degrading the prefetch. 
Instead, \name  allows the indices of later slices to be 
packed with those of the current slice by marking the first index of the next slice with a {\em start} bit, which is the $6^{th}$ metadata bit out of 7 in~\secref{representation}.  The indices in the normal value-index column-read also use the {\em start} bit to indicate the  first index within a slice.

SDDS continues to handle the uncertainties in the decoupling so that \name remains headless with little on-chip control despite the decoupling. Recall that SDDS packs the non-zero cells into a compressed matrix (\secref{SDDS}). 
When the next non-zero index is in a  later vector slice, SDDS inserts an invalid  cell as before and additionally sets the cell's {\em start} bit. When a valid cell is the first index within the corresponding slice, then also the SDDS sets the cell's {\em start} bit. Additionally,  SDDS also simulates the iFIFOs so that if any of the banks' iFIFO  is full then SDDS places a {\em placeholder} index in the matrix column which the full iFIFO drops during execution. The former invalid index (no match in the corresponding slice)
enters the  iFIFO whereas the  latter placeholder (no room in the iFIFO) does not. 
The iFIFO holds the index, {\em invalid} and {\em start} bits in each entry.

Two cases are possible for the  entries at the heads of the iFIFOs in a bank: (1) If any of the {\em start} bits is false or an iFIFO is empty implying that some cells in a later column-read may match some vector elements from the current slice, then SDDS stalls the  vector broadcast while the current slice stays latched. In the stalled 
broadcast time slot, the iFIFO entries with the {\em start} bits set to false extract the matching vector elements from the latched slice  into the corresponding eFIFOs, after which the iFIFOs are advanced. The iFIFO entries with the {\em start} bits set to true do not affect the eFIFO; those iFIFOs are not advanced. 
(2) If all the iFIFO head entries' {\em start} bits are true, then the next 
vector broadcast occurs (a different command than a broadcast stall), as directed by SDDS. All the valid iFIFO entries extract
the matching elements from the broadcast into the corresponding eFIFOs, any invalid iFIFO entry does not insert any element into
the eFIFO, and  the iFIFOs are advanced. Invalid indices imply zero values which SDDS does not place in the compressed matrix, mirroring no element being inserted into the eFIFO.
SDDS stalls the broadcast if an eFIFO is full.  SDDS records these  stalls in the command stream generated for the host (\secref{SDDS}). 

In either case above, irrespective of whether an insertion occurs into the eFIFO (tail), which stays ahead of the matrix values due to the decoupling, a column-read triggers multiplication of the 
values from the column-read and vector elements at the heads of the eFIFOs.  The column-read values are consumed immediately and the eFIFOs are advanced. In the rare case that an eFIFO is empty (i.e., the vector element matching the matrix value in the column-read is delayed), SDDS places a zero matrix value in the compressed matrix.

\putsubsec{switch}{Simplifying the switch}

Instead of the brute-force 16x11 switch to select the vector elements from the broadcast matching the matrix cell indices in the column-read, we exploit the $t_{CCD}$-constrained time between the vector broadcasts to simplify to a 4x11 switch that 
can be used sequentially four times per broadcast ($t_{CCD}$ is usually 4). This switch is simply 11 4-to-1  multiplexers, one per execution unit (\figref{switch.png}). In cycle $i$ ($0 \leq i \leq 3$), the multiplexer extracts an element for the eFIFO if the  iFIFO head index falls in the range 4$i$ to 4$i$+3. While each 4-to-1 multiplexers uses its iFIFO's lower-order two index bits for select, the upper two index bits are compared to the constants 0, 1, 2, and 3 to 
determine if the index is in the desired range. 
The input to the switch itself chooses the $i^{th}$ among four sets of four contiguous sub-vector elements at indices 4$i$ to 4$i$+3 (\figref{switch.png} left). Thus, there is at most one iFIFO read and at most one  eFIFO write each cycle (\secref{decouple}).  Because the MACs compute different inner products in our layout,  the same
element may match more than one eFIFO entry. Also, some iFIFOs  
may have invalid indices and may not match any element.
An alternative simplification to a 16x3 switch time-shared by 11 MACs over 4 cycles can select at most one element per MAC in 4 cycles whereas the 4x11 switch can select over 4 cycles more than one element per eFIFO (those falling in different index ranges) when the iFIFO has more than one index.

\figput{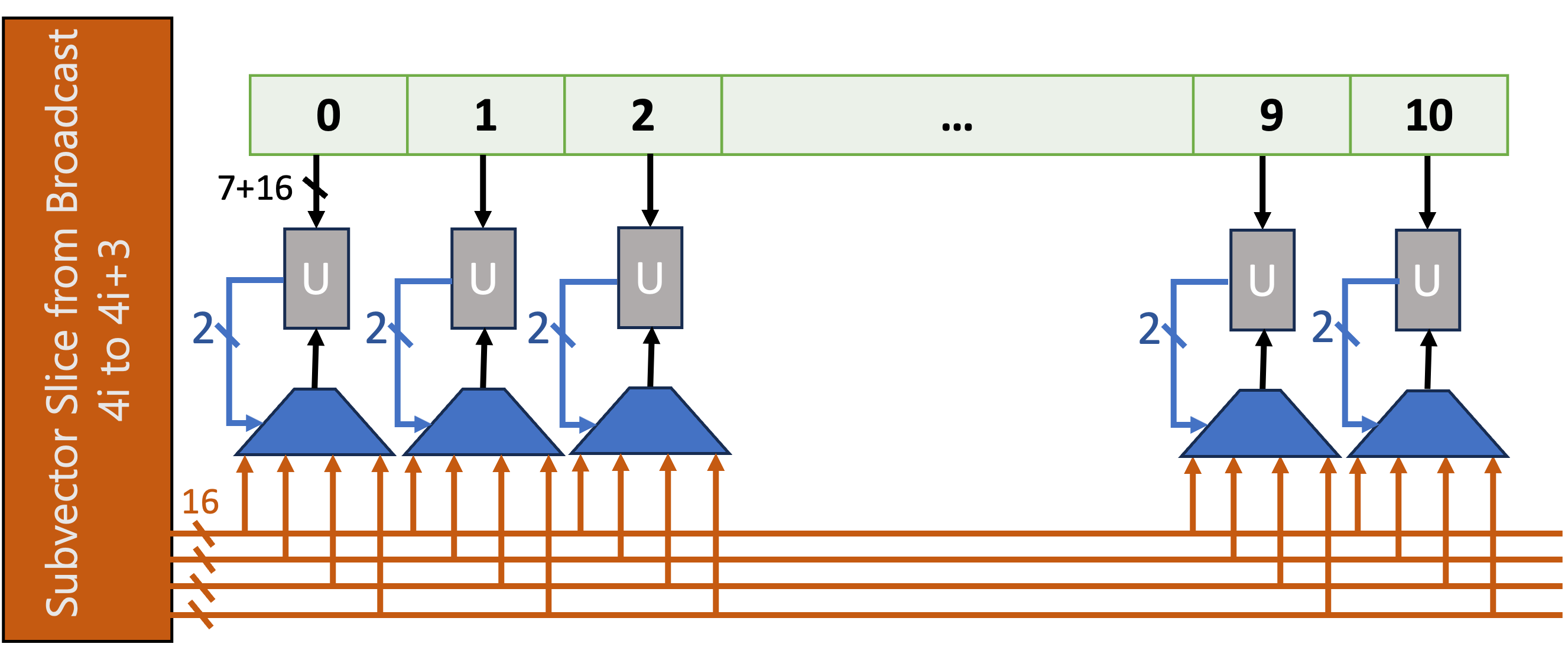}{}{\name's simplified switch}

Consecutive indices in an iFIFO belonging to the same range cannot be handled in the same broadcast (consecutive cycles of
the same broadcast handle different ranges). This condition forces broadcast stalls because the iFIFO is a strictly in-order FIFO. Instead, reordering the indices and the corresponding matrix cells, such that consecutive indices within the same matrix column read are from different ranges, avoids most stalls.  For example, assume the index ranges are 0-3, 4-7, 8-11, and 12-15, and  indices {\em i2}, {\em i3}, {\em i5} and  {\em i6} are in an iFIFO.  
In the first range, {\em i2} is consumed but {\em i3} forces a broadcast stall. Further, though {\em i5} is in a different range than {\em i3}, {\em i5}
cannot be consumed because of head-of-line blocking by {\em i3}. Thus, these indices need a broadcast (handles {\em i2}) and two stalls: the first stall handles {\em i3} and {\em i5} and the  second stall {\em i6}. However, reordering the indices (and their corresponding cell values) as {\em i2}, {\em i5}, {\em i3}, and {\em i6} results in one broadcast (for {\em i2} and {\em i5}) and only one stall (for {\em i3} and {\em i6}). 
SDDS performs this reordering to improve performance and inserts the necessary broadcast stalls for correctness.

\putsubsec{gb}{Load balance}

A remaining issue is load imbalance across sparse and dense rows in different banks that happen to be processed synchronously, causing MAC idling in the banks with the sparse rows. 
\name adopts SparTen's greedy load balancing~\cite{sparten} which sorts the matrix rows by density and assigns the sorted rows to the banks in a round-robin fashion, while co-locating 
within each bank the  densest row and the  sparsest, the next densest and sparsest rows, and so on. In this co-location, the dense and sparse rows are intermingled in logically-increasing index order. Our fine-grained interleaving is applied after the co-location.
To ensure that each
matrix cell contributes to the correct output element, we add a {\em select} bit per matrix cell which is the $7^{th}$ metadata bit out of 7 in~\secref{representation}. 
Accordingly, each  bank has two output buffers. 

\putsubsec{other}{Other issues}
ML models employ activation functions after most layers (not to be confused with DRAM row activation). Because there are many choices for these functions which are changed by ML practitioners, \name like Newton offloads the functions to the host. 
The host can apply simple functions, such as ReLU, hidden under the result read-out~\cite{Newton}. However, more complex functions that need to scan the result vector, such as softmax, cannot be hidden easily but can be vectorized on the host. We account for this overhead in our results. Finally, because PIM does not check ECC, which occurs in the host memory controllers, \name adopts Newton's assumption of periodically reloading the matrix~\cite{Newton}. The unused ECC bits (32 per 256 data bits in HBM2~\cite{hbm2e1,hbm2e2}) can allow adding another MAC per bank in \name.

\figput{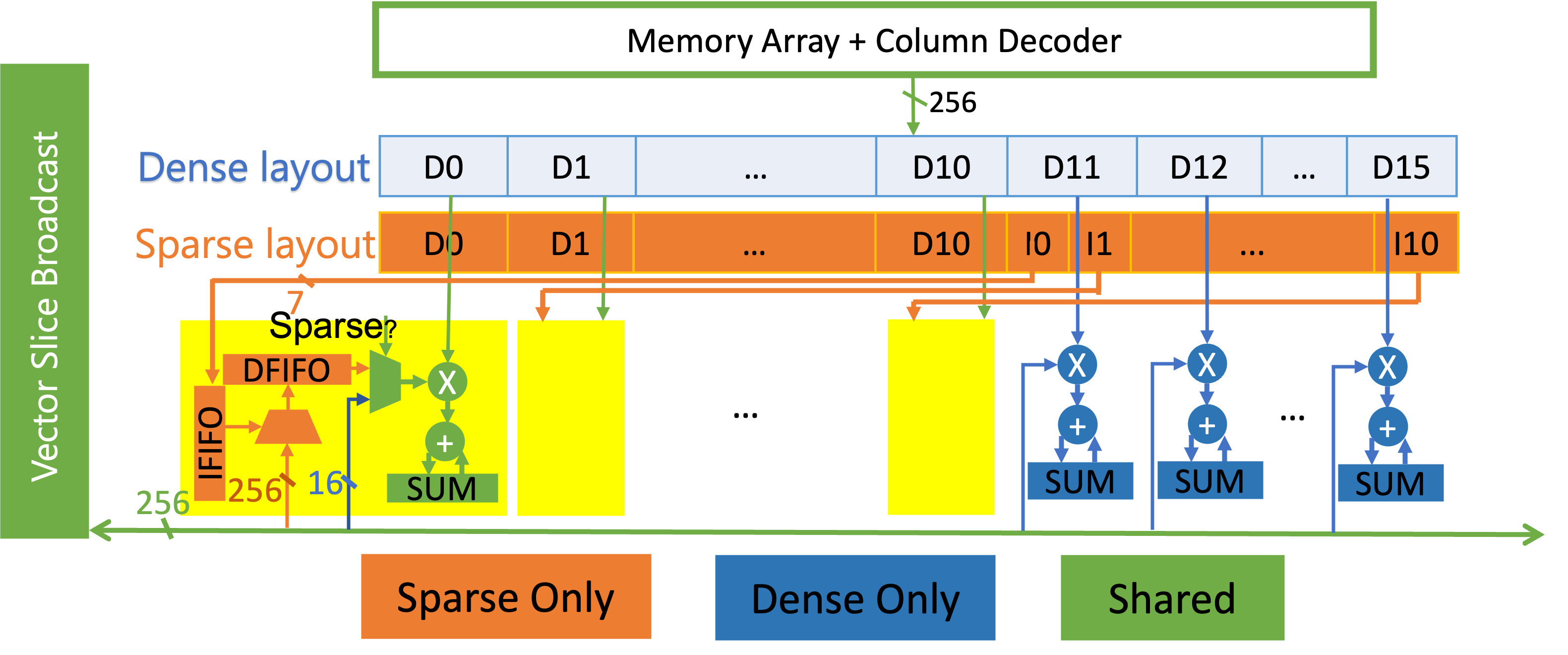}{}{Flexible configuration for both sparse and dense models}

\putsubsec{flexible}{Flexibly supporting both dense and sparse models}

While pruning is a well-established technique for generating sparse models, some dense models may continue to be used because pruning does take effort. As such, we extend \name to
support flexibly both sparse and dense models. Recall that the dense models need 16 MACs for 16 dense matrix elements per column (\figref{Newton_datapath.png}) whereas the sparse models need 11 MACs (\figref{datapath.png}).  In the extension, the same MACs can be used in either case except that the 11 
MACs for the sparse models are accompanied by the FIFOs and the switch (shown to the left in~\figref{flexible.png}) while the remaining 5 MACs for the dense models do not have the FIFOs (shown to the right in~\figref{flexible.png}).  Accordingly, the sparse index metadata needs to be laid out carefully to avoid 
excessive multiplexing in the datapath. To that end, for the sparse matrices we place the 11 matrix elements contiguously followed by their index metadata in the same order. In~\figref{flexible.png}, the dense layout simply
shows the 16 elements {\em D0} through {\em D15}
whereas the sparse layout shows the 11 elements 
{\em D0} through {\em D10} followed by {\em I0} through {\em I10}.
Thus, irrespective of sparse or dense matrix, the first 11 elements' positions are identical ({\em D0} through {\em D10} in~\figref{flexible.png}). In the case of dense matrices, the next 5 elements follow whereas for sparse matrices the index metadata follows ({\em I0} through {\em I10}). 
This layout efficiently achieves this flexible support with the only extra hardware of 2-to-1 multiplexers for the vector input to the MACs to choose between the vector broadcast data (for dense models) and the eFIFO output (for sparse models). 
To avoid energy overhead, the FIFOs and switch needed for the sparse models are power-gated off during dense model inference, and the last 5 MACs are power-gated off during  sparse model inference.

\tabput{commands}{
\begin{tabular}{ll}
\hline
\bf Command & \bf Operation\\ \hline
LOAD-GB\# & Load global buffer chunk\# \\
ALL-ACT & All-bank activation \\
LOAD-IDX\# & Load column\# into iFIFOs \\
 & (no broadcast or compute) \\
COMP-NoBR\# & Compute column\# with eFIFO, load iFIFO, \\ 
 & extract  from stalled vector slice into eFIFO \\
COMP-BR\# & Compute column\# with eFIFO, load iFIFO, \\ 
 & extract  from broadcast vector slice into eFIFO \\
RDRES\# & Read result vector of bank\# 
\end{tabular}
}{Key commands in \name}

\putsec{method}{Methodology}

\tabput{hw-param}{
\begin{tabular}{|rl|}
\hline
Num of Ranks  & 1 \\
Num of Banks &  16   \\ 
Num of Rows in each bank & 32768\\
Num of Column I/Os per row &  32 \\
Column I/O bit width & 256b (16 \bfloat)\\
Num. of MACs per bank &  11 (16 used in dense) \\
Num. of entries per FIFO per MAC & 8 \\
\hline
\multicolumn{2}{|c|}{\bf Timing Parameters (in DRAM Cycles)}\\
\multicolumn{2}{|c|}{$t_{RAS}=24$; $t_{RCD}=10$; $t_{RRD}=4$; $t_{RC}=34$}\\
\multicolumn{2}{|c|}{$t_{RP}=10$; $t_{CCD}=4$; $t_{RTP} = 5$; $t_{WTR}=5$}\\
\hline
\end{tabular}
}{DRAM Configuration (HBM2E-like)}

\noindent
{\bf PIM simulation: }
Based on DRAMsim2~\cite{DRAMSim}, our cycle-level simulator
captures the key details of \name's commands (\tabref{commands}).
The basic DRAM parameters  (e.g., banks, row/column widths)  are  similar to HBM2E's ({\tabref{hw-param}}).
Our DRAM configuration uses an 8-high stack with 8 channels, 2 pseudo channels, and 16 
banks per channel for a total capacity of 128 Gb. Each bank 
has 32K rows and 8K columns. Each row of 8K bits (or 1K bytes) can be accessed at a 256-bit 
column I/O granularity to which ESPIM's MACs per bank are rate-matched(16 for dense MV and 11 for sparse MV).  
The simulator models refresh. We compare to Newton with 16 MACs per bank (for uncompressed sparse matrix with no index overhead) and to SpaceA ~\cite{spacea} with  a 4-KB CAM, 512-entry associatively-searched 
load queue, and a 2-KB scratchpad per bank. 
SpaceA's area estimates show 10\% overhead over DRAM assuming only one MAC per bank which does not saturate the bank bandwidth.
We estimate SpaceA's area using CACTI~\cite{cacti} to arrive at 3 MACs per bank for equal area as \name.

\noindent
{\bf Non-PIM architectures: }
To compare \name against non-PIM architectures, 
we consider an ideal non-PIM host, \inpim, which models an upper-bound on performance of {\em any} non-PIM architecture including processing-near-memory (PNM) proposals (e.g.,~\cite{Rajeev-PNM,Chrystos-PNM,Mutlu-PNM,nda, TETRIS,neurocube})
and traditional systems (GPU, TPU, and multicores).
Assuming unlimited compute resources, \inpim is limited only by the DRAM's external bandwidth so that  \inpim's 
execution time is only the data transfer time between the DRAM and  host.  
\name's speedups against realistic non-PIM architectures, including multicores, GPUs, TPUs, or any custom non-PIM (PNM or traditional) accelerator, would only be  higher than \name's speedup  over \inpim.

\noindent
{\bf GPU simulation: }
We use GPGPUsim\cite{gpgpusim} (version 4.0), to model a realistic, high-performance non-PIM host (as opposed to the unrealistic \inpim discussed above). 
We configure GPGPUsim  as a Titan X,  a high-end model with 3072 CUDA cores and 24 memory channels. On the software front, we use  Cutlass-1.3 \cite{gpgpusim-cutlass, cutlass}, a high-performance, open-source CUDA library for linear algebra. 
Cutlass incurs a 
large  constant time overhead that hurts the GPU's performance, as reported by Newton~\cite{Newton}. Following Newton, we eliminate this overhead by running several matrix-vector computations to isolate the incremental 
cost of each matrix-vector computation. This elimination {\em reduces} the GPU's execution time; including any part of the overheads would make the GPU only worse.

All the architectures use the same DRAM parameters.

\tabput{dataset}{
\begin{stripetabular}{ccc}
\bf Workload &  Matrix & Vector \\ 
\hline
Attention.wk  & $4096 \times 4096$ & $4096 \times 1$ \\
Attention.wo  & $4096 \times 4096$ & $4096 \times 1$  \\
Attention.wq  & $4096 \times 4096$ & $4096 \times 1$  \\
Attention.wv  & $4096 \times 4096$ & $4096 \times 1$  \\
Feed\_forward.w1 & $11008 \times 4096$ & $4096 \times 1$  \\
Feed\_forward.w2 & $4096 \times 11008$ & $11008 \times 1$ \\
Feed\_forward.w3 & $11008 \times 4096$ & $4096 \times 1$ \\
\hline
\end{stripetabular}
} {Benchmarks}

\noindent
{\bf Benchmarks:}
We use \textit{Large Language Model Meta AI} (LLaMA-7B)\cite{llama}, whose size fits edge deployment, pruned to various sparsities. The sizes of LLaMA's various matrices are shown in \tabref{dataset}.  LLaMA-7B employs 30 modules each of which has 4 attention layers and 3 feed-forward layers. We run all these layers. As discussed in~\secref{other}, we offload the activation functions to the CPU whose overhead we include and isolate in our results.
The rest (e.g., normalization and embedding) is 0.092\% of time for LLaMA-7B{~\cite{overhead}}.
Previous work~\cite{cerebras, sparsegpt} has reported achieving 80-90\% sparsities while maintaining accuracy. However, the pruned models are not available publicly. 
Therefore, we prune the model by choosing the pruning thresholds as per the pruning algorithm~\cite{pruning} to achieve various reported sparsities.  Such pruning leads to unstructured sparsity. Because we do not study the pruned models' accuracies  which are reported elsewhere~\cite{cerebras, sparsegpt}, we do not perform time-consuming retraining which recovers accuracy without changing sparsity.

\noindent
{\bf Energy and area: }
PIM (Newton and \name) and GPU differ in energy as follows: (1) While GPU does not incur compute energy in the DRAM, PIM's compute in each bank consumes about 4 times the energy of DRAM reading consecutive columns from the same DRAM row~\cite{Newton}. We conservatively ignore GPU's 
compute energy which is hard to estimate without a detailed energy model (GPU’s energy can only be worse). (2) GPU incurs energy to transfer the matrix whereas PIM incurs transfer energy for the input vector and the partial results which are far smaller than the matrix. (3) \name's dummy cells inserted by SDDS incur energy overhead which do not exist in Newton and  GPU.
For area, Newton's MACs incur about 25\% area over conventional DRAM~\cite{Newton}. However, detailed area and energy for logic in DRAM process is not publicly known. Instead, we implement {{\name's}} datapath in Verilog (MACs, FIFOs, latches, and the switch) and synthesize at 45-nm technology using  FreePDK45{~\cite{pdk}}. Because FreePDK45 does not include an SRAM library and our FIFOs are too small for CACTI, we use flip flops for our FIFOs which incur much larger area and energy than SRAM (so the FIFO area and energy are likely to be better).

Using the MACs' area factor of 25\% for Newton, we scale the area of \name's full datapath. Similarly,  using the above compute energy factor of 4x for Newton, we scale the energy of \name's full datapath. 

Our SDDS implementation takes about 10 minutes  on a 16 Cores of Intel E5-2623 to schedule our benchmarks. Finally, we measure vectorized softmax runtime on Intel E5-2623 with 4 cores per channel to add as overhead to \name.

\figputW{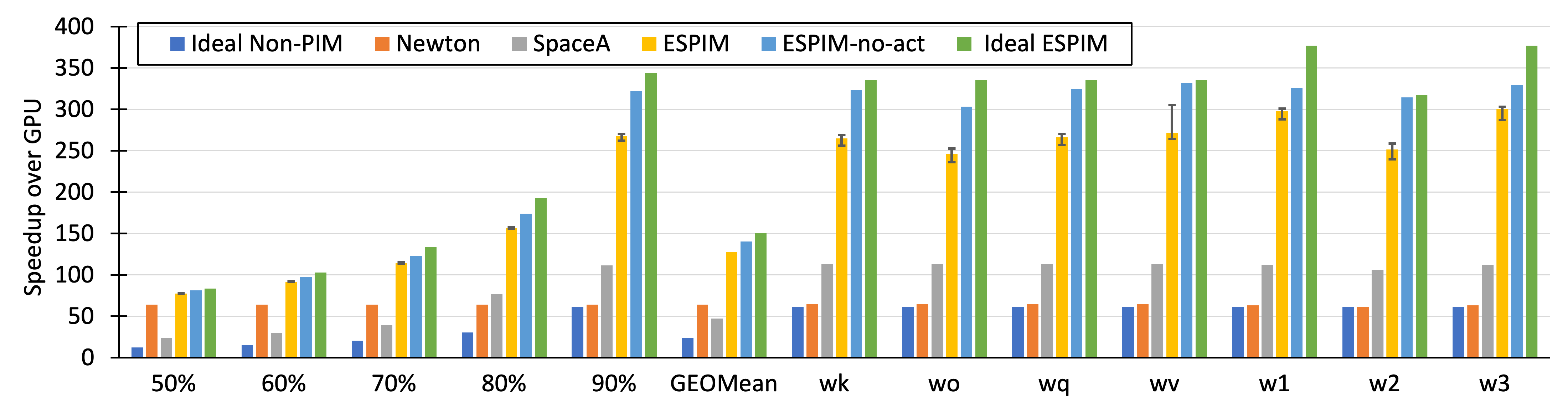}{}{Speedup}

\putsec{results}{Results}

We start by comparing the performance of \name and other architectures. We also isolate the performance impact of \name's techniques. We then show \name's sensitivity to the FIFO sizes and the number of banks. Next, we compare \name and Newton in terms of energy. Finally, we compare the area overhead of  \name and Newton.

\putsubsec{perf}{Performance}

\figref{speedup.png} shows the speedups of \inpim, Newton, SpaceA, \name, \name without ML activation function (\name-no-act) and \iespim over a Titan X-like GPU 
for (1) the full model with the sparsity varied as 50-90\% in steps of 10\% 
and (2) individual layers of the model at 90\% sparsity (X-axis). 
The full model runs include activations (\secref{other}), whereas the individual layers do not. We also show \iespim which is an ideal version of \name without any stalls. While the GPU and Newton use uncompressed sparse matrices, all the others use compressed sparse matrices.
All the architectures except \inpim and \name-no-act incur ML activation function overhead (\inpim's unlimited compute eliminates this overhead).
For \name, we show the range of the speedups (not standard deviation) across (a) all the layers of the full model and (b) all the instances of each model layer (\secref{method}).

Because \inpim's execution time is limited only by DRAM-host data transfers (\secref{method}), \inpim's speedup improves with more sparsity (to the right) as less data is transferred between the DRAM and host. However, \inpim's limited  speedup, 28x on average, motivates PIM in general. Being a dense PIM which does not exploit sparsity, Newton's speedups do not change with sparsity. At high sparsities (e.g., 90\%), \inpim despite being pin-bound catches up with Newton by exploiting sparsity. Due to its limited compute (3 MACs per bank), SpaceA performs worse than Newton at low sparsities and then improves (at 90\% sparsity, Newton effectively has only 10\% of the bank bandwidth and 1.6 MACs). 
\name performs better than both \inpim (PIM effect) and Newton (sparsity effect), achieving 127x mean speedup over GPU (2x over Newton). In the time for one external DRAM row transfer by \inpim, PIM (Newton and \name)  can consume a DRAM row in each bank. The gap between \name-no-act and \name, which shows the ML activation overhead, increases with sparsity as the MV computation is sped up more. \iespim adds to \name's speedup by avoiding \name's stalls though 
at lower sparsities (to the left) \name is closer to \iespim due to less
irregularity and fewer stalls.  
\iespim achieves lower than perfect, sparsity-implied speedups over Newton (e.g., at 90\% sparsity \iespim achieves around 5.6x speedup over Newton instead of 6.9x) due to Amdahl's Law limit imposed by ML activation. Finally, the individual layers (to the right) show similar trends as the full model at 90\% sparsity though there is diversity among the layers.  \name's speedups have little to modest variance across layers.

\name performs nearly identically to Newton for dense
models (not shown) because the number of MACs, and vector input and result output traffic are the same for Newton and \name.

\putsubsec{isolate}{Isolating individual optimizations}
To isolate the impact of \name's optimizations, \figref{isolation.png} shows  \name's speedup over the GPU for the benchmarks (X-axis) as we {\em progressively} add the optimizations {\em one at a time} leading up to full \name. 
We start with the fine-grained interleaving without which performance is poor. We add decoupled prefetch, reordering to alleviate the simplified switch's conflicts, and greedy balancing. We also show the large 16x11 switch to isolate the impact of our simplification. Even at low sparsities (to the left),  where \name's opportunity and irregularity in the computation are low, \name's decoupled prefetch boosts performance. As the opportunity and irregularity increase with more sparsity, \name's reordering and greedy balancing contribute more, especially at 90\% sparsity. 
Finally, there is little gap between \name and the large switch even at high sparsities, confirming the soundness of our decision to simplify.

\figput{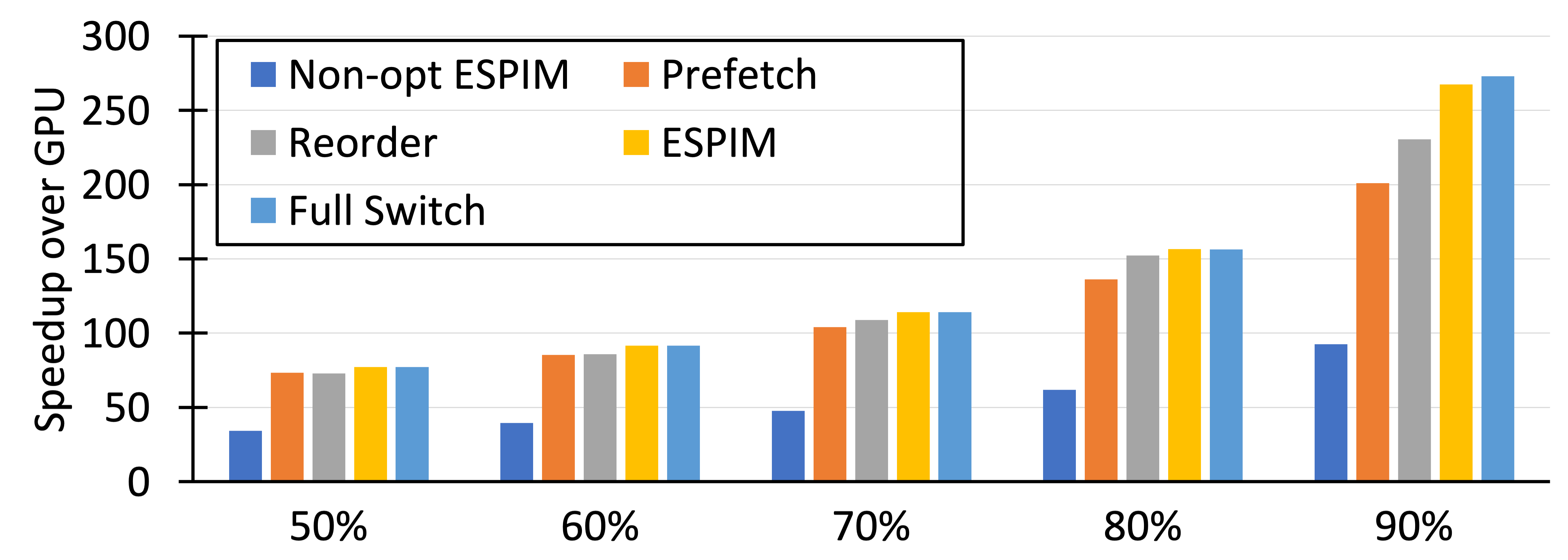}{}{Isolating \name's optimizations}

\putsubsec{fifosensitivity}{Sensitivity to FIFO size}

\figref{FIFOsize.png} shows the speedup of \name over the GPU (Y-axis) for the 
benchmarks (groups of bars on the X-axis) as \name's iFIFO and eFIFO sizes are varied (individual bars in each group). As expected, \name's speedups at a given sparsity improve with longer FIFOs which absorb more irregularity. More sparsity results in more irregularity (from left to right), so that longer FIFOs provide more improvements.

\figput{FIFOsize.png}{}{Sensitivity to FIFO size}

\putsubsec{banksensitivity}{Sensitivity to Number of Banks}

\figref{bank.png} shows the speedup of \name over the GPU (Y-axis) for our 
benchmarks (groups of bars on the X-axis) as  the number of banks is varied (individual bars in each group).
Because the compute and memory bandwidths increase proportionally with number of banks, \name's speedups increase with more banks. However, with more sparsity (from left to right), the higher irregularity dampens this  speedup growth as does the DRAM row activation overhead but to a lesser extent because the activation overhead is low due to all-bank activation.

\figput{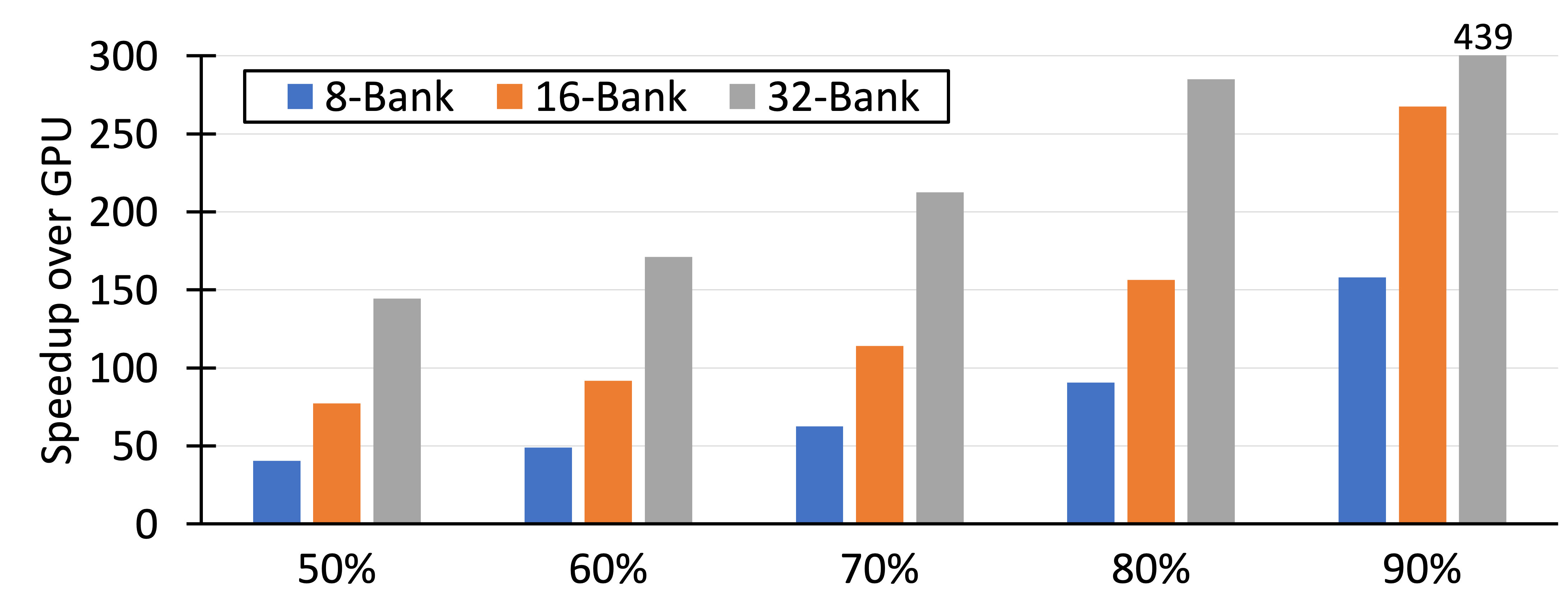}{}{Sensitivity to number of banks}

\putsubsec{energy}{Energy}
\figref{energy.png} shows energy normalized to that of GPU's conventional DRAM (Y axis) for the benchmarks (X axis). 
PIM (Newton and \name) energy includes compute whereas any 
non-PIM architecture (multicores or GPUs) would  incur compute energy and host-memory transfer energy in addition to memory energy which are not included in the DRAM energy. As such,
\name's energy is likely to remain lower. We break down energy into {\em access}, {\em compute} and {\em rest} (only for \name's extra hardware).
Though Newton uses uncompressed matrices, we assume that Newton gates the MACs for zero values to save energy. However,
Newton incurs the access energy for the full uncompressed matrix. Newton's dense matrix energy overhead of around 1.8x is almost entirely due to its compute; this overhead reduces with sparsity due to the MAC-gating. Assuming the flexible configuration for sparse and dense  models (\secref{flexible}), \name incurs only slightly more overhead than Newton for the dense matrix because the FIFOs needed for sparse models are power-gated off (only the small 2-to-1 multiplexers for the vector input to the MACs are extra).
For the sparse matrices,  \name dissipates lower energy than
Newton by capturing sparsity even in the access unlike Newton. However, ESPIM incurs sparsity-related overheads not in Newton, including the indices, FIFOs and switch (shown as {\em rest}), which decrease with increasing sparsity. Note that this overhead is conservative given our implementation uses bulky flip flops and multiplexers for the FIFOs (\secref{method}), instead of efficient SRAM. More so, the access overhead of the sparse representation pushes ESPIM's  energy  above the sparsity-proportional fraction of Newton's energy at full density. 
For instance, ESPIM's energy at 50\% sparsity (1.8x) is higher than half of Newton's at full density (2.8x).
Nevertheless, \name's 2x higher performance and 34\% lower energy than Newton illustrate \name's energy efficiency.

\figput{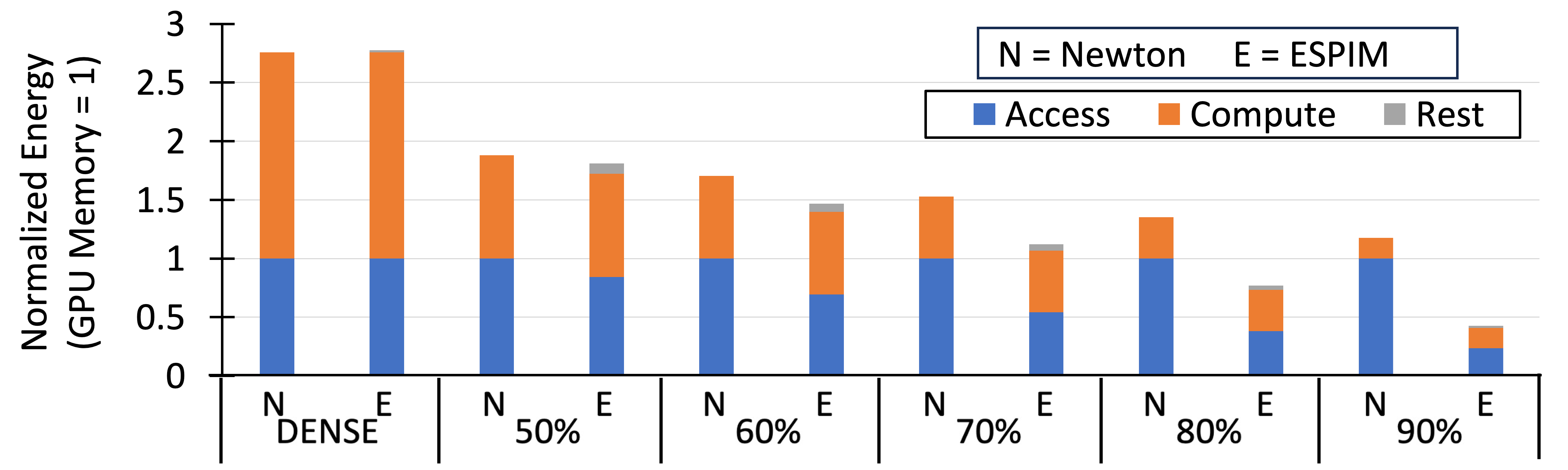}{}{Energy}

\tabputW{area}{
\begin{stripetabular}{ccccc}
\hline
\bf  Newton &  Norm. area  & Description  & &\\ 
\hline
Newton MACs & 25\%  & 16 MACs   & &\\
\hline
\hline
\bf ESPIM & Sparse-only norm. area & Description & Sparse+dense norm. area & Description\\
\hline
ESPIM MACs & 17.2\% & 11 MACs & 25\% & 16 MACs\\   
ESPIM iFIFO & 3.5\% &  11 8X7b FIFO & 3.5\% & 11 8X7b FIFO\\
ESPIM eFIFO & 7.1\% & 11 8X16b FIFO & 7.1\% &  11 8X16b FIFO\\
ESPIM Switch + other logic & 3.0\% & 11 16b 4-1 Mux + other logic & 4.1\% & 11 16b 4-1 Mux + other logic \\
\hline
ESPIM Total & 30.8\% & Sparse-only ESPIM & 39.7\% &  Sparse+dense ESPIM\\
\hline
\end{stripetabular}
} {Area}

\putsubsec{area}{Area}

While Newton incurs 25\% area over conventional DRAM~\cite{Newton}, \name's components and their area for  the configuration supporting only sparse models {\figref{datapath.png}} and  the flexible configuration supporting both sparse and dense models (\figref{flexible.png}).
are listed in~\tabref{area}. Because of its fewer MACs than Newton in its sparse-only configuration, ESPIM recovers some area which is spent on the FIFOs and switch. In total, the area overhead for ESPIM's sparse-only configuration is around 31\% over conventional DRAM and under 5\% over Newton.
In return, this configuration achieves 5.4x and 2x speedups  for sparse models over \inpim and Newton, respectively.
Because of using the same number of MACs as Newton and extra 2-to-1 multiplexing  for the vector data input to the MACs (counted in ``other logic'' in~\tabref{area}), the flexible configuration's area overhead increases to  under 40\% over conventional DRAM and under 12\% over Newton. As discussed above, our flip flop-based FIFO implementation makes these area  overheads also conservative.

\putsec{related}{Related work}

PIM and PNM have a long history as the idea has been revisited multiple times over several decades in the context of various technologies (e.g., analog versus digital PIM), architectures (e.g., general-purpose, versus SIMD) and workloads (e.g., general purpose, graph analytics, and map reduce)~\cite{Kogge-PIM,Kogge-DIVA,Chrystos-PNM,Rajeev-PNM,IRAM,IBM-AMC,xsd,Chrystos-PNM, millipede,Mutlu-PNM,pim-Pipelayer,pim-ISAAC,pim-Prime,AC-DIMM,reorg,mcdram}.
Recent PIM proposals from DRAM vendors, Function-In-Memory (FIM)~\cite{fimdram, FIM-ISSSC} and Accelerator-in-Memory (AiM)~\cite{Newton, Newton-ISSSC} target MV computation, a key kernel for ML inference (especially transformer-based models). 
We have discussed Newton in detail. In contrast to Newton's headless architecture, FIM employs programmable cores for generality at the cost of area and power. While we have described
\name based on Newton, FIM's datapath is similar to Newton's.
Further, FIM can also benefit from sparsity's  energy and performance advantages.  
As such, \name's techniques are applicable to FIM as well. 

SpaceA is a sparse PIM that targets hyper-sparse MV for HPC. As extensively discussed, SpaceA takes a hardware-intensive approach to combat such extreme sparsities. To each bank, SpaceA adds a scratchpad for the matrix,
a CAM for the vector, an associatively-searched queue to extract the matching vector elements, and independent control to handle sparsity's uncertainty and irregularity.
Instead, for moderate sparsities in ML, \name uses the DRAM row buffer to hold the matrix column, broadcasts the vector slices by exploiting the DRAM's organization and decouples the indices and values to hide the sequential broadcast delays, uses  a simplified switch to extract the matching vector elements, and  employs SDDS to continue to be a headless architecture and avoid much on-chip control.

Other, non-PIM sparse architectures target sparse MM  in ML~\cite{cnvlutin,scnn,sparten,dstc,cambricon,cambriconx, s2ta} and  hyper-sparse MM in HPC~\cite{outerspace,sparch,spaghetti,matraptor,gamma}.

\putsec{concl}{Conclusion} 

PIM promises to improve the performance and energy of memory pin-bandwidth-bound matrix-vector (MV) computation in prevalent ML inference. These improvements can be amplified by unstructured sparsity in ML models. Thus, our target is unstructured, one-sided, weight-only sparsity where the vector is dense. However, PIM imposes stringent constraints on area and energy whereas unstructured sparsity introduces uncertainty, irregularity and load imbalance in PIM's all-bank synchronous operation. \name addresses these challenges via four contributions. First, because matrix sparsity 
increases the  vector broadcast bandwidth demand for every matrix column-read, 
\name reduces the demand by sharing each vector broadcast among multiple rows in each bank
via  a {\em fine-grained interleaving} of the matrix cells.
Second,  to remain  a {\em headless}, datapath-only architecture which  mostly avoids on-chip control's area and energy
despite sparsity's uncertainties, \name  exploits  the observation that the sparsity is data-dependent  but  static and known at training. 
Accordingly, \name employs {\em static data-dependent scheduling (SDDS)} to derive the sparse MV's cycle-level schedule and to  insert the appropriate stalls for correctness. 
Third, to alleviate any long delay between a matrix cell's column-read 
and the broadcast of the matching vector element,  places the indices  ahead of the matrix cell values, {\em decoupling the indices and values} 
to enable prefetching of the vector elements. We extend SDDS for performance and correctness with the decoupled prefetching.

Finally,  we {\em simplify the switch} required to select the vector elements that match the matrix cells instead of a brute-force, impractically-large design. 
We extend SDDS to improve performance by achieving fewer conflicts in the simplified switch. Our simulations showed that \name achieves 2x average (up to 4.2x)
speedup over and 34\% average (up to 63\%) lower energy
than Newton while incurring under 5\% area.
These results make a compelling case for sparse PIM architectures targeting emerging sparse ML models that are pin-bound.   
\newpage

\bibliographystyle{IEEEtranS}
\bibliography{paper}

\end{document}